\documentclass[twocolumn,showpacs,showkeys,preprintnumbers,superscriptaddress,amsmath,floatfix,amssymb,secnumarabic,nofootinbib]{revtex4-1}
\usepackage[colorlinks=true]{hyperref}
\usepackage{graphicx}
\usepackage{dblfloatfix}    
\usepackage{placeins}
\usepackage{braket}
\usepackage{amsmath}
\usepackage{epsfig}
\usepackage{float}
\usepackage{nicefrac}
\usepackage{xcolor}
\usepackage{color}
\usepackage{bm}
\usepackage{subfigure}
\usepackage{chngcntr}
\usepackage{ulem}

\def\({\left(}
\def\){\right)}
\def\[{\left[}
\def\]{\right]}

\newcommand{\ve}[1]{\boldsymbol{#1}}

\newcommand{\diag}[1]{ {\rm diag} \, \left( #1 \right) }
\newcommand{\Tr}{ {\rm Tr} \, }

\newcommand{\beq} {\begin{eqnarray}}
\newcommand{\eeq} {\end{eqnarray}}

\newcommand{\comment}[1]{}

\begin{document}
\sloppy

\title{Beyond the instanton gas approach: dominant thimbles approximation for the Hubbard model}

\author{Maksim~Ulybyshev}
\email{maksim.ulybyshev@uni-wuerzburg.de}
\affiliation{Institut f\"ur Theoretische Physik und Astrophysik, Universit\"at W\"urzburg, 97074 W\"urzburg, Germany}

\author{Fakher~F.~Assaad}
\email{fakher.assaad@uni-wuerzburg.de}
\affiliation{Institut f\"ur Theoretische Physik und Astrophysik, Universit\"at W\"urzburg, 97074 W\"urzburg, Germany}
\affiliation{W\"urzburg-Dresden Cluster of Excellence ct.qmat, Am Hubland, 97074 W\"urzburg, Germany}

\begin{abstract}
   To  each complex  saddle point  of  an  action,  one  can attach  a Lefschetz  thimble on which  the imaginary part of the 
action is  constant.  Cauchy  theorem  states  that   summation over  a set of  thimbles  produces  the  exact  result.    This  reorganization
of  the path  integral,  is an  appealing  starting  point  for  various approximations:  In the  realm of auxiliary  quantum Monte 
Carlo  methods  it provides a  framework to  alleviate  the  negative sign problem. 
Here, we  suggest  to  constrain the  integration to  the  \textit{dominant} thimbles:  the thimbles  attached to the  saddle  points with  the 
largest statistical  weight.  For  the  Hubbard model,  in a  formulation where the  the  Hubbard Stratonovitch field  couples to the  charge,  
this  provides a  \textit{symmetry} consistent   approximation to the physics  of the  Hubbard  model:  constraining the integration domain 
does not  explicitly break a  symmetry.  We can test this  approach  for 
the  Hubbard model at half-filling on a bipartite lattice.  The  paper  builds  on the previously developed instanton 
gas approach, where an exhaustive saddle point approximation was constructed.  
We  present  results,  showing  that the   dominant  thimbles approximation provides  results  that are in 
remarkable   agreement with the exact results for various fermionic 
observables including spin and charge order parameters and single electron spectral functions. We discuss implications of our results 
for simulations away of half filling. 
\end{abstract}

\maketitle


\section{\label{sec:Intro}Introduction}

Quantum Monte  Carlo (QMC) methods   for   frustrated spin systems  or for  models of  correlated  electrons  generically  
suffer   from   the so called  negative  sign  problem \cite{Troyer:2004ge}.  Here,  stochastically  sampled configurations 
carry  a  sign such  that  destructive  interference results in  a  total  average sign  that  vanishes   exponentially  with  system size 
 and   inverse  temperature.   Owing to  the  central limit  theorem,  the  uncertainty  in the measurement  scales  as  
 the  inverse  square  root  of the  computational  resources.  
For    bounded  computational  resources this leads  to  fluctuations   that  quickly  run out  of  control.   This  cancellation is   formulation 
dependent  and  considerable  research has   been  carried  out  so as  to alleviate this problem \cite{Hangleiter20,WanZQ22}.   In fact,  
optimal  formulations  have  the potential  of reaching  energy  scales  and  system  sizes  that   can  impact  
understanding  of  non-trivial experimental  relevant  phenomena \cite{SatoT21,Huang19}.   

In the  realm  of  the   auxiliary   field   quantum Monte  Carlo  approach,   a  possible  route  to  alleviate the 
negative  sign  problem   is  the so called  Lefschetz thimbles   approach  \cite{Witten:2010cx,Witten:2010zr,Cristoforetti:2012su}.   
Let  $\mathbb{R}^N$  be  the  integration  
domain of  the fields $\ve{\Phi}$  and $S$  the  action  such  that the   partition  function  reads:
\begin{equation}
   Z  = \int  d \ve{\Phi}  e^{- S(\ve{\Phi})}. 
\end{equation}
Cauchy theorem allows the deformation of the integration contour without changing the final value of the integral. 
Let  us hence  complexify the  fields $\Phi$    and  search  for  the saddle point of  the  complexified  action. 
To   each  saddle point,  one  can  attach  a  manifold  in $\mathbb{R}^N$  that  is  coined 
a  Lefschetz thimble.  Importantly,   on the   Lefschetz  thimble,  the  imaginary  part  of the action is  
constant such that   configurations belonging to  the  same  thimble  do not  cancel out providing the fluctuations of complex measure are not too severe.   
However, since  one  has  to  sum  over  many  thimbles,  it  may  very  well  be  that  the contributions of  different  thimbles 
cancel out  such  that  ultimately no progress  is  achieved.

   The  question  we  pose  in  this  article  is if  this  cancellation  does  not  occur   for non-trivial  cases. 
In  fact   we  wish  to  inquire  if   constraining  the  integration over  a  set of \textit{dominant}  thimbles with fixed $\operatorname{Im} S$   is an accurate enough approximation to the full path integral.   Generically,  this  is a  daunting task:   we have  to  find the  dominant   saddle points,  
constrain  the integration  over  the  corresponding   dominant  thimbles  and   have  benchmark results at  hand,  so as  to  gauge  the 
quality  of the  approximation.

This  program  can  be  achieved  for  the  special  case of  the  Hubbard model at half-filling on a bipartite lattice. 
Particle-hole   symmetry  allows  us  to formulate  a  negative sign  free    algorithm such that  numerically   exact  
results  on finite lattices  are    at  hand in  polynomial   time.

Due  to  the  absence of  the  negative sign  problem  thimbles   are  confined to the  real  domain,  and   are  separated  by  the  
zeros  of  the  fermion  determinant, or  equivalently  by logarithmic  divergences in the  action.   In  fact,  the   weight of  the 
configuration  is   proportional to the  square  of  a  fermion determinant,  that  by  symmetry  turns out  to  be  real.   It 
is  worth  noting  that  the  sign  of  a   single   determinant  fluctuates   very  strongly  such  that  a  full  Monte Carlo  simulation  will  sample  many   thimbles.  
In previous papers \cite{PhysRevD.101.014508, PhysRevB.107.045143} we could describe the general structure of the saddle points 
for the Hubbard model depending on the employed Hubbard-Stratonovich transformation. The structure is particularly simple, when the 
auxiliary field  couples  to  the  charge density operator. In this case the saddle points can be easily described by the so called 
instanton gas model \cite{PhysRevB.107.045143}, which gives us an accurate prediction of the dominant saddles: the one, where  
the corresponding  thimbles contributes  at most  to the partition function. Based on this, we wrote an exhaustive saddle 
point approximation for the Hubbard model at half filling. This model accounted  for  local moment  formation  and  demonstrated remarkable agreement with unbiased QMC simulations 
in the structure of the high energy part of the spectrum of the single particle excitations. However, the pure saddle point approximation 
did not provide a correct spectrum in the vicinity of the Fermi level: it missed the appearance of the mass gap and the 
formation of the  Mott insulator. 

Here we  go one step  beyond the saddle point approximation.  Using  QMC techniques confined within a set of designated thimbles, 
and the previous predictions concerning the structure of the saddle points for the Hubbard model, we show that  integration over
the  dominant thimbles accounts  completely  for  the physics of the Hubbard model at least at half-filling:   spectral functions 
and order parameters are   next  to  indistinguishable from the exact  results.  

At present,  we are able to confirm the validity of this approximation only  at half filling, when we avoid the difficulties 
of sampling the fields over the thimble in complex space, corresponding  to  integration over a a complicated curved manifold. 
There are however, strong indications that the structure of the thimble decomposition for the Hubbard model remains the same away 
of half filling provided  that  the auxiliary field  couples  to  the charge \cite{PhysRevD.101.014508}. This differs from previous papers where a  single-thimble approximation is  considered   in the  realm of  a  decomposition where  the 
auxiliary field  couples to the magnetization  \cite{Cristoforetti}.  This   suggests  that the dominant thimbles approximation may  also  be  a good  approximation  away from half-filling. 
In this case it will substantially alleviate the sign problem, since the phase of the action is constant by construction, 
and only the fluctuations of the complex measure appearing due to integration over curved manifold in complex plane contribute 
to the sign problem. In was shown however, that the sign problem arising  from the fluctuation of  the measure  is   substantially weaker than  the one arising  from the phase of the action  \cite{Alexandru:2016ejd,PhysRevD.101.014508}.    Hence we can conjecture  that  QMC on the dominant thimbles can 
boost our ability to simulate the  Hubbard model away of half filling at  lower temperatures and on larger lattice sizes than 
 present methods. 

The paper is organized as follows. In Sec.~\ref{sec:formalism_Lefschetz}  we  give  a    brief  summary   of 
 Lefschetz thimbles and of  the  instanton gas formalism for the Hubbard model. 
  Sec.~\ref{sec:formalism_QMC} gives an account of the Monte Carlo process bounded within the designated set of thimbles.  Here  we  
  equally  demonstrate   that  this  approximation  yields  surprisingly  good  results  
 for   spin and charge order parameters as well as  for  the   single-particle spectral functions.
 Finally  in  Sec.~\ref{sec:Conclusion} we summarize our results and discuss further perspectives.

  \begin{figure}[]
   \centering
   \subfigure[]{\label{fig:HistogramExample}\includegraphics[width=0.35\textwidth,clip]{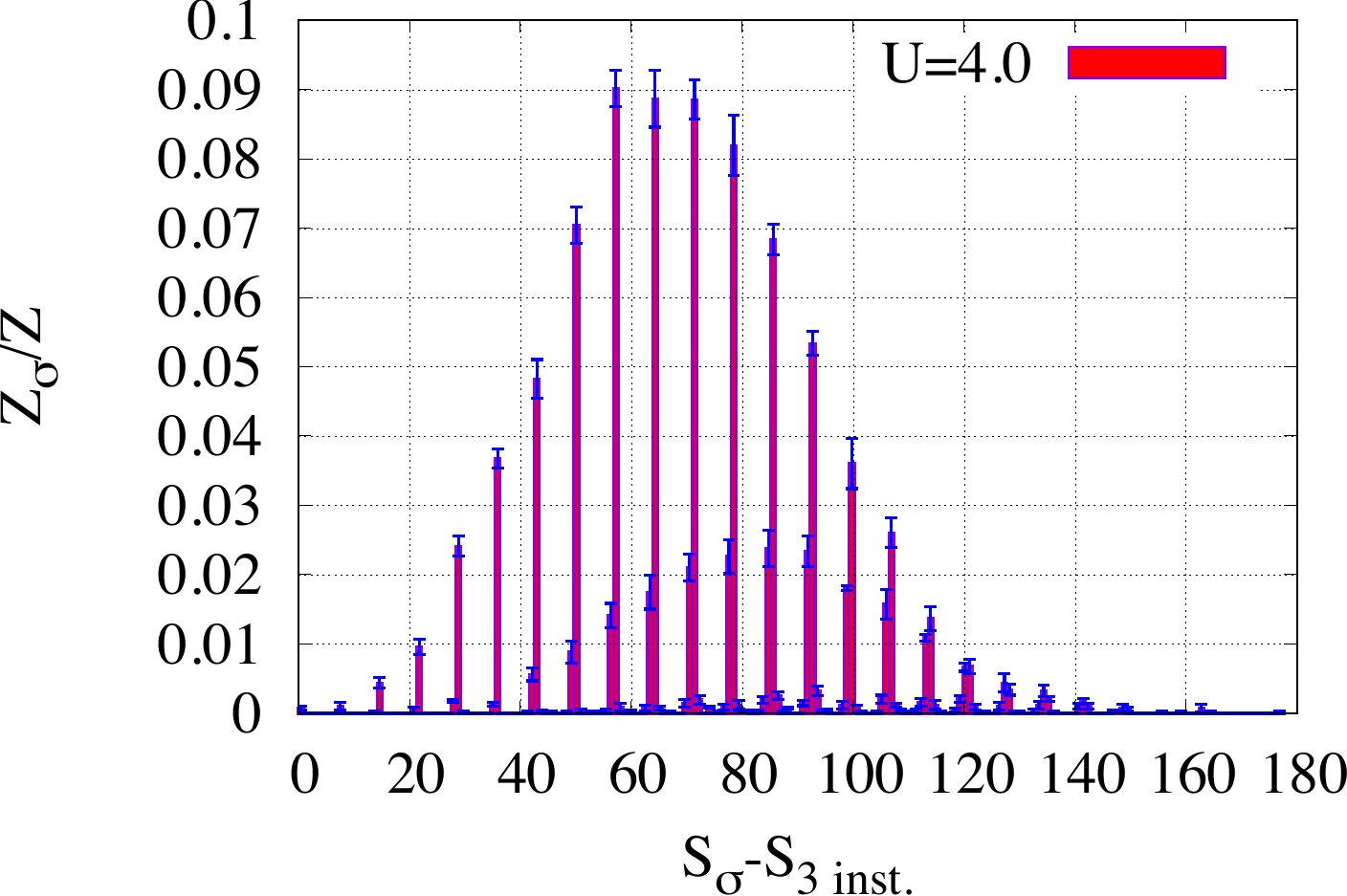}}
   \subfigure[]{\label{fig:DensityComparison}\includegraphics[width=0.35\textwidth,clip]{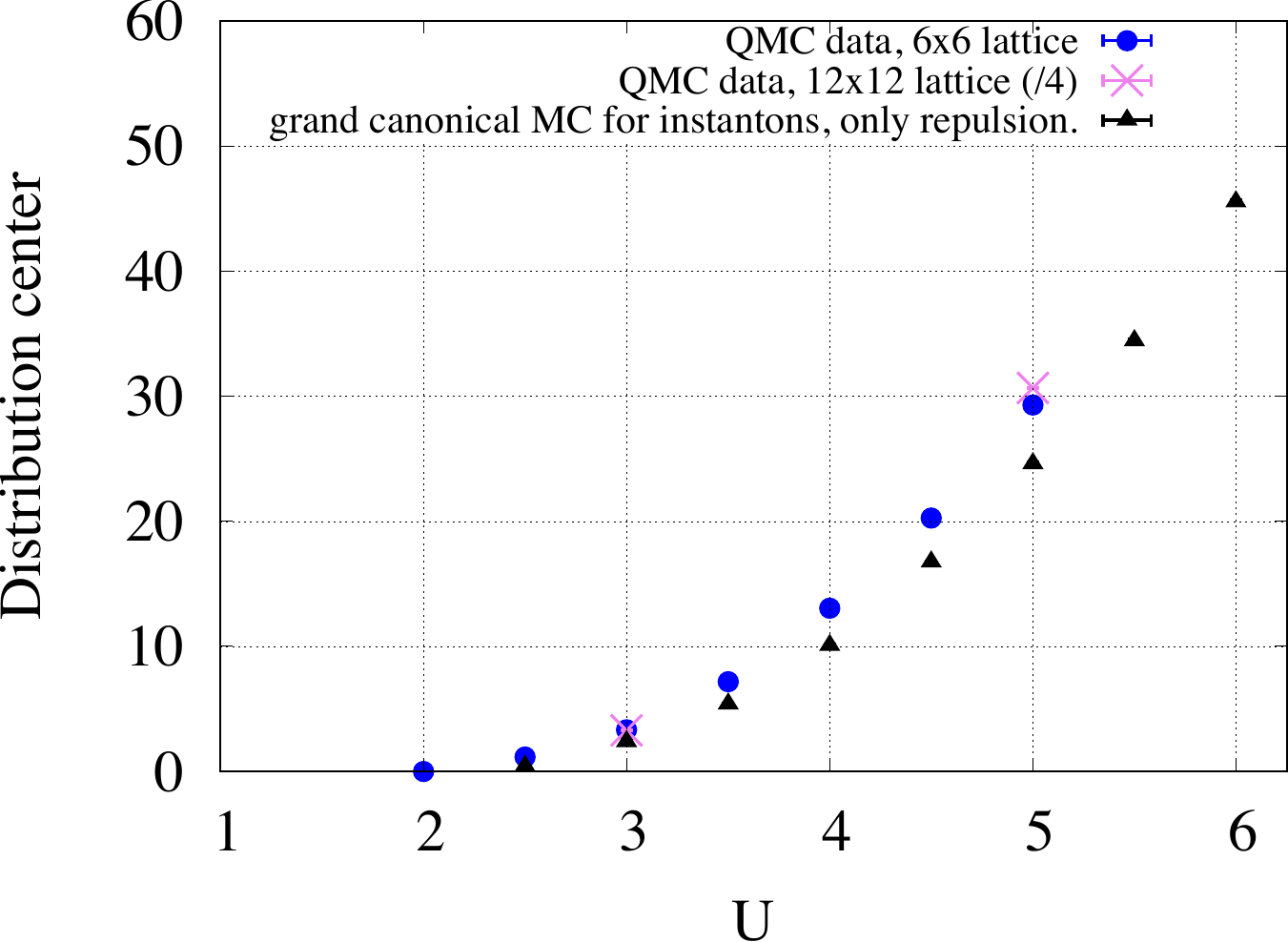}}
      \caption{(a) Histogram showing the distribution of the actions of the QMC configurations after the downwards gradient flow. The flow is long enough for the stabilization of the field configurations at the saddle points. The discrete distribution   that results  in   \textit{quantized}  values of  the action  $S$ for  a  saddle point 
      field configuration, $\Phi_\sigma$, 
      is the manifestation of the instanton gas \cite{PhysRevB.107.045143}. Calculations are done on the $6\times6$ lattice with $N_t=512$ and $\beta=20$. The action is plotted with respect to the action of 3-instantons saddle point. (b) Number of instantons plus anti instantons corresponding to the maximum in the distribution of the instanton number (density of instantons) depending on the interaction strength. We compare the QMC results obtained on $6\times6$ and $12\times12$ lattices (note the rescaling of the $12\times12$ lattice data to bring it to the same scale) with the prediction of the instanton gas model with hardcore repulsion between instantons. $N_t$ and $\beta$ are the same as for the plot (a). }
   \label{fig:InstantonGasConstruction}
\end{figure}

\section{\label{sec:formalism_Lefschetz}Lefschetz thimbles formalism and instanton gas approach for the Hubbard model}

\subsection{\label{subsec:formalism_intro} Lefschetz thimbles formalism for the Hubbard model}

The Hubbard model on a bipartite  lattice can be written in particle-hole basis as:
\begin{eqnarray}
\hat H = \hat H_{tb} + \hat H_{\mu} +\hat H_U= \nonumber \\-\kappa \sum_{<{\ve{x}},{\ve{y}}>} (  \hat a^\dag_{\ve{x}} \hat a_{\ve{y}} + \hat b^\dag_{\ve{x}} \hat b_{\ve{y}}  + \mbox{h.c}) + \mu \sum_{\ve{x}}q_{\ve{x}} + \frac{U}{2} \sum_{\ve{x}} \hat q^2_{\ve{x}},
\label{eq:ham}
\end{eqnarray}
where $\hat a_{\ve{x}}$ and $\hat b_{\ve{x}}$ are annihilation operators for electrons and holes, 
and $\hat q_{\ve{x}}=\hat a^\dag_{\ve{x}} \hat a_{\ve{x}} - \hat b^\dag_{\ve{x}} \hat b_{\ve{x}} $ 
is excess electrical charge  at site ${\ve{x}}$  with  respect  to the half-filling  condition. 
The kinetic energy includes hoppings between sites belonging to different sublattices. In the following, 
all dimensional parameters including interaction strength $U$ are expressed 
in  units of the hopping $\kappa$. We included the chemical potential $\mu$ to consider some general properties of the partition function and the thimble decomposition, however, $\mu$ is set to zero in practical simulations.

We are interested in the path integral for the partition function 
\begin{eqnarray}
\mathcal{Z}=\Tr \left( e^{-\beta \hat H} \right),
\label{eq:z_tr}
\end{eqnarray}
where $\beta$ is the inverse temperature. After the introduction of the charge coupled continuous auxiliary field,
the partition function can be written as follows  \cite{Assaad_complex, Buividovich:2018yar}:
\begin{eqnarray}
 \mathcal{Z} = & \int \mathcal{D} \phi e^ {-\int_0^\beta d \tau \sum_{{\ve{x}}}  \frac  {\phi_{\ve{x}} (\tau)^2 } {2 U}} \Tr \left(  e^{-\int_0^\beta d \tau \hat H(\hat{a}^{\dagger},\hat{a}^{},\hat{b}^{\dagger},\hat{b}^{}, \phi(\tau) ) } \right),  \nonumber \\
  & \hat H(\hat{a}^{\dagger},\hat{a}^{},\hat{b}^{\dagger},\hat{b}^{}, \phi (\tau)) = H_{tb} + H_{\mu} + i \sum_{{\ve{x}}} \phi_{\ve{x}}(\tau) \hat q_{\ve{x}}.
  \label{eq:Z_general_cont}
\end{eqnarray}  
After Trotter decomposition and appropriate rescaling of the $\phi$ field, we finally get
 \begin{eqnarray}
  \mathcal{Z} = \int \mathcal{D} \phi e^ {-S_B[\phi]}  \det M_{el.}[\phi] \det M_{h.}[\phi], \nonumber \\
   S_B[\phi] = \sum_{{\ve{x}},\tau}  \frac  {\phi_{{\ve{x}},\tau}^2 } {2 U \Delta \tau}.
  \label{eq:Z_general}
\end{eqnarray}  
$\Delta \tau$ is the step in Euclidean time, $N_\tau \Delta \tau = \beta$. 
Auxiliary field $\phi_{{\ve{x}},\tau}$ is introduced for each site $\ve{x}$ in each
 Euclidean time slice $\tau$. $M_{el.}$ and $M_{h.}$ are the fermionic operators for the electrons
  and holes respectively. The determinants of these operators are expressed as
\begin{eqnarray}
 \det M_{el.} = \det \left[ I +\prod^{N_\tau}_{\tau=1} D_{2\tau-1} D^{el.}_{2\tau} \right], \nonumber \\
 \det M_{h.} = \det \left[ I +\prod^{N_\tau}_{\tau=1} D_{2\tau-1} D^{h.}_{2\tau}\right], 
  \label{eq:M_continuous}
\end{eqnarray}
where $D^{el.}_{2\tau} \equiv \diag{e^{i \operatorname{Re}\phi_{{\ve{x}},\tau}  -\operatorname{Im}\phi_{{\ve{x}},\tau}+\mu \Delta \tau} }$, $D^{h.}_{2\tau} \equiv \diag{e^{-i \operatorname{Re}\phi_{{\ve{x}},\tau}  +\operatorname{Im}\phi_{{\ve{x}},\tau}-\mu \Delta \tau} }$ and $D_{2\tau+1} \equiv e^{-\Delta \tau h_{tb}}$ have
 been introduced (we included the possibility for the complexification of the fields $\phi$).
Both of these are $N_S \times N_S$ matrices, where $N_S$ is the total number of spatial lattice sites. 
We have also introduced $h$, which is the matrix characterizing the tight-binding part of the Hamiltonian 
$\hat{H}$. The determinant $\det M_{el.}$ ($\det M_{h.}$)  is  a  result  of  the integration  over  the 
fermionic particle (hole)  field  $\hat a$ ($\hat b$ ).

The integral \ref{eq:Z_general}  can be rewritten in general form 
\begin{eqnarray}
\mathcal{Z} = \int D \Phi e^{-S(\Phi)}, 
\label{eq:z_action}
\end{eqnarray}
where $\Phi=\{ \phi_{x, \tau} \}$ is the set of all auxiliary fields, and the action includes both 
Gaussian part and logarithms of the fermionic determinants:
\begin{eqnarray}
S = S_B(\Phi) - \ln \det M_{el.} - \ln \det M_{h.}. 
\label{eq:action}
\end{eqnarray}  

The Lefschetz thimbles formalism \cite{Witten:2010zr} is a way to rewrite the general integral 
\ref{eq:z_action} using the information on the saddle points of the action $S$.
The initial integral \ref{eq:z_action} is written for real auxiliary fields $\Phi$. 
However, the exponent $e^{-S}$ in \ref{eq:z_action} is analytical function 
of the fields $\Phi$. Thus we can change the integration contour shifting it into complex plane 
without changing the final result for the partition function. 

  \begin{figure}[]
   \centering
   \subfigure[]{\label{fig:actions}\includegraphics[width=0.35\textwidth,clip]{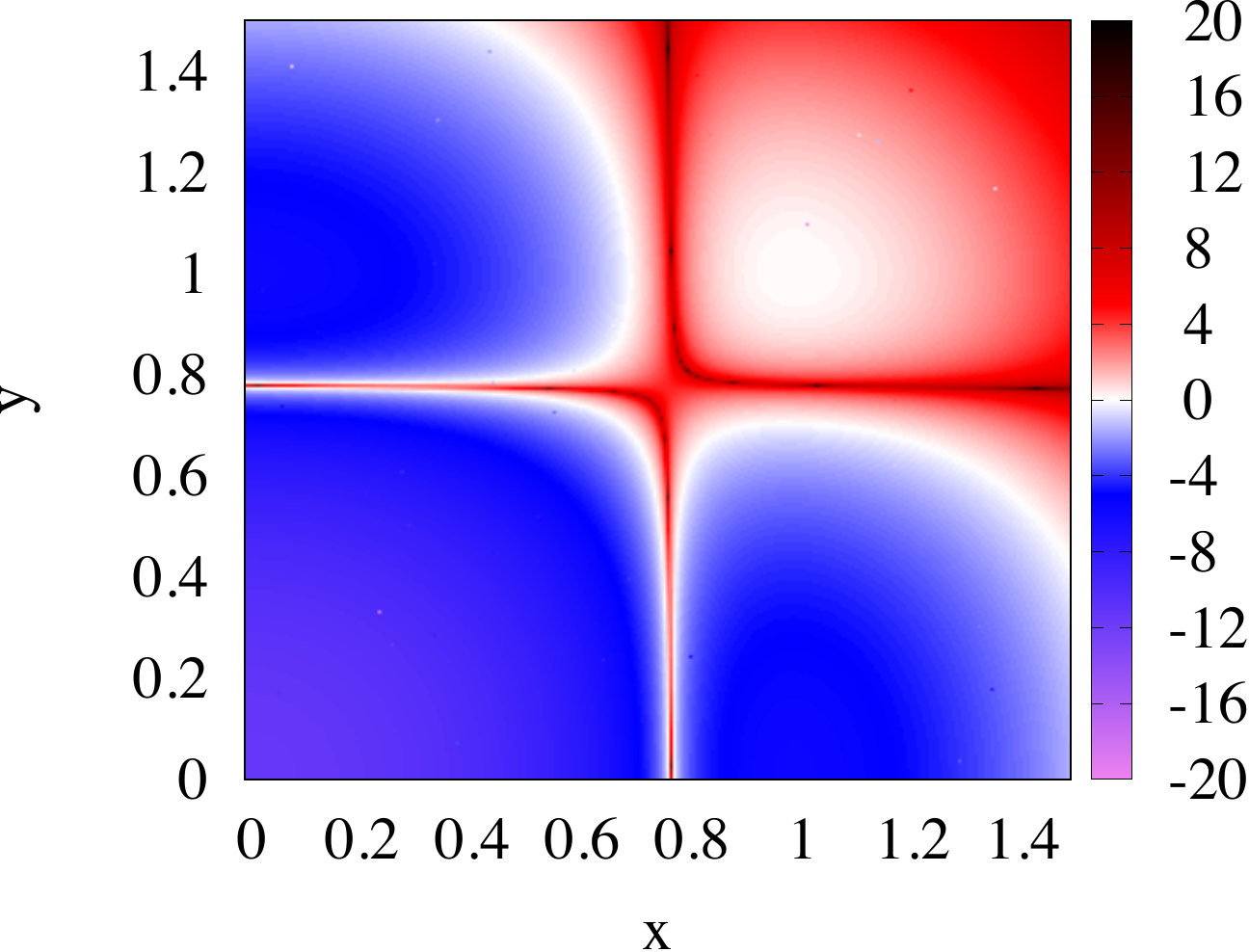}}
   \subfigure[]{\label{fig:spins}\includegraphics[width=0.35\textwidth,clip]{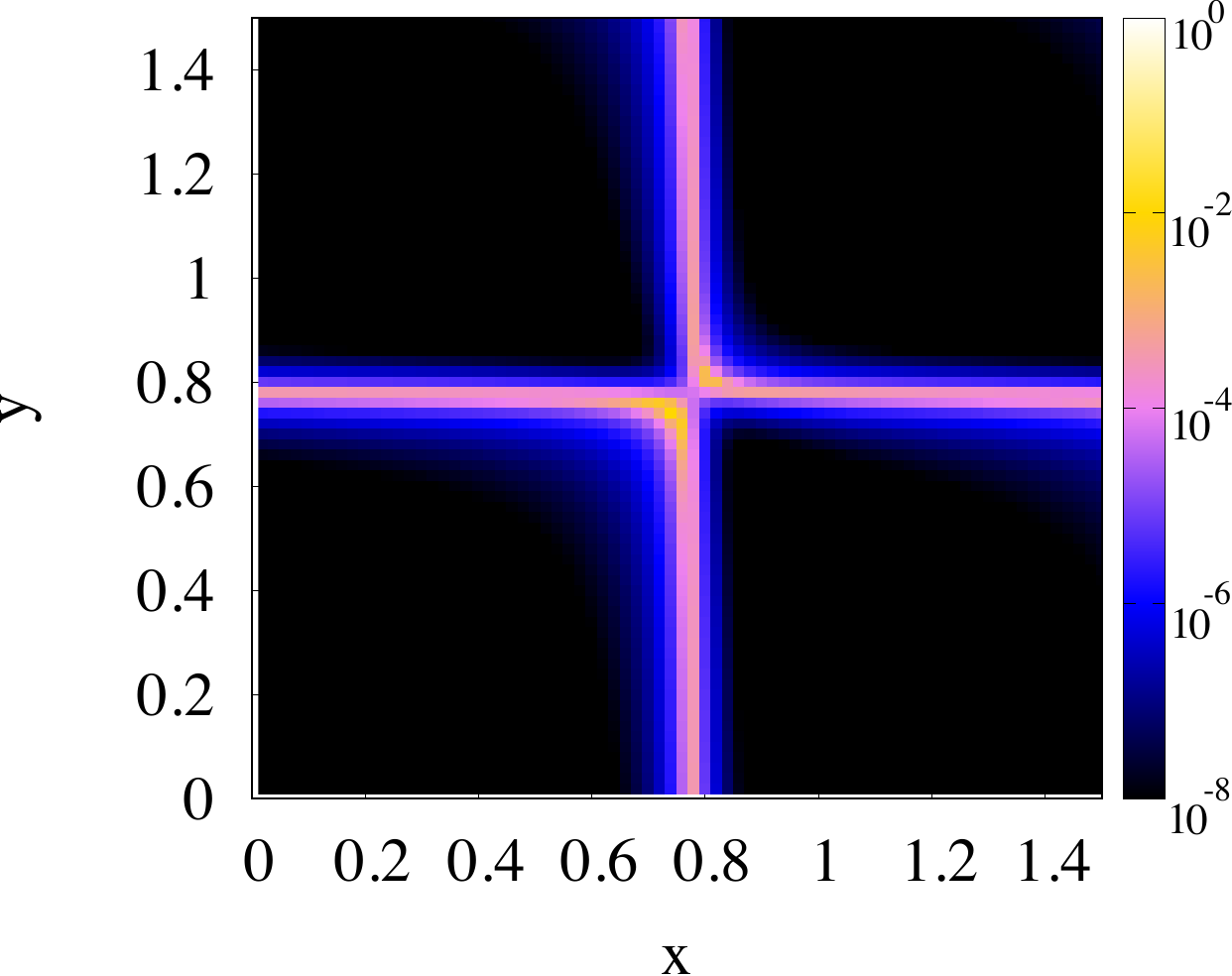}}
      \caption{(a) Action of the field  $\phi  =  x \phi^{(1)} + y \phi^{(2)} $   where $ \phi^{(i)} $ are 
      single  instanton configurations at  different  space  time  positions. 
        Thus (1,1) point is two-instanton saddle, (0,1) and (1,0) are different one-instanton saddles and (0,0) is the  vacuum.
         The  distance between  the   two instantons is set  to  3 lattice  constants in real  space and $\delta \tau=10$ in  Euclidean time.
         The color scale is linear with zero corresponding to the action of the two-instanton saddle point. 
         (b) The ratio of the long range spin-spin correlator to the short-range spin-spin correlator \ref{eq:spin_ratio}. 
          Note the logarithmic color scale. Computations for both plots were carried  out  on a $6\times6$ lattice at $\beta=20$  and  interaction strength $U=6$.}
   \label{fig:ActionsSpins}
\end{figure}

Let $\Phi_\sigma$ be the saddle points of the action $S$. Note that the saddle points can be complex.
 Each saddle point features two manifolds emanating from it:  the thimble, $\mathcal{I}_\sigma$, and 
 the anti-thimble,  $\mathcal{K}_\sigma$. 
Thimble $\mathcal{I}_\sigma$ is a union of all solutions of the Gradient Flow (GF) equations
\begin{eqnarray}
\frac{d\Phi}{d t}=\overline{ \frac{\partial S}{\partial \Phi}},
\label{eq:gr_fl}
\end{eqnarray}
which start at the saddle point at flow time $t=-\infty$: 
$\Phi(t) \in\mathcal{I}_\sigma$ if  $\Phi( t \rightarrow -\infty) \rightarrow \Phi_\sigma$.
The  anti-thimble $\mathcal{K}_\sigma$ consists of all possible    solutions of the GF $\Phi(t)$
 which end up at a given saddle point $\Phi_\sigma$:  $\Phi(t) \in\mathcal{K}_\sigma  $  if
  $ \Phi(t \rightarrow + \infty) \rightarrow \Phi_\sigma$.  We also need the number of 
  intersections of an anti-thimble with real subspace: 
  $k_\sigma = \langle \mathcal{K}_\sigma, \mathbb{R}^N \rangle$. 
Both thimble and anti-thimble are $N$ dimensional real manifolds in complex 
space $\mathbb{C}^N$, where $N=N_\tau N_S$ is the total number of auxiliary fields.

  \begin{figure}[]
   \centering
   \subfigure[]{\label{fig:HistogramU3Before}\includegraphics[width=0.23\textwidth,clip]{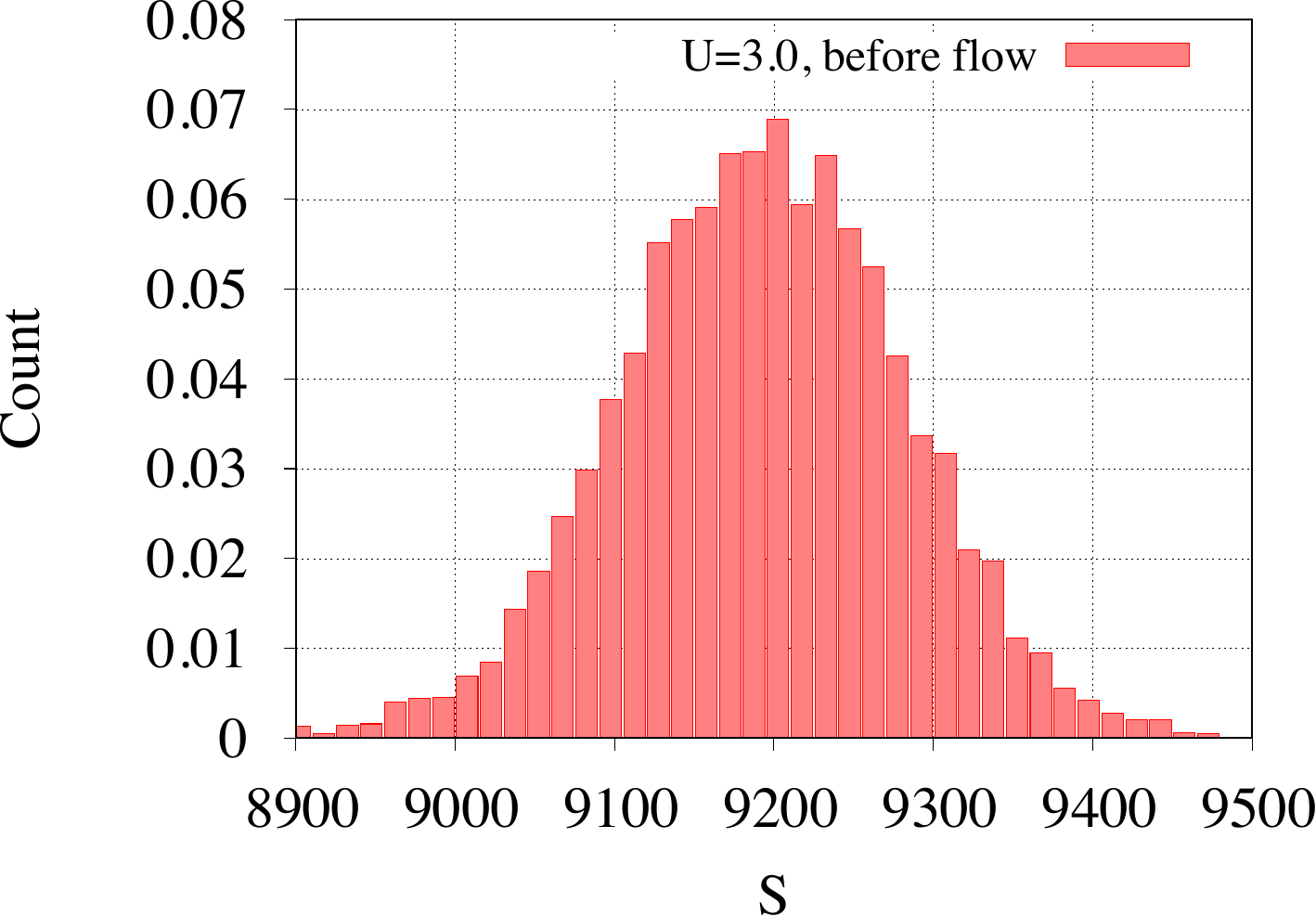}}
   \subfigure[]{\label{fig:HistogramU4Before}\includegraphics[width=0.23\textwidth,clip]{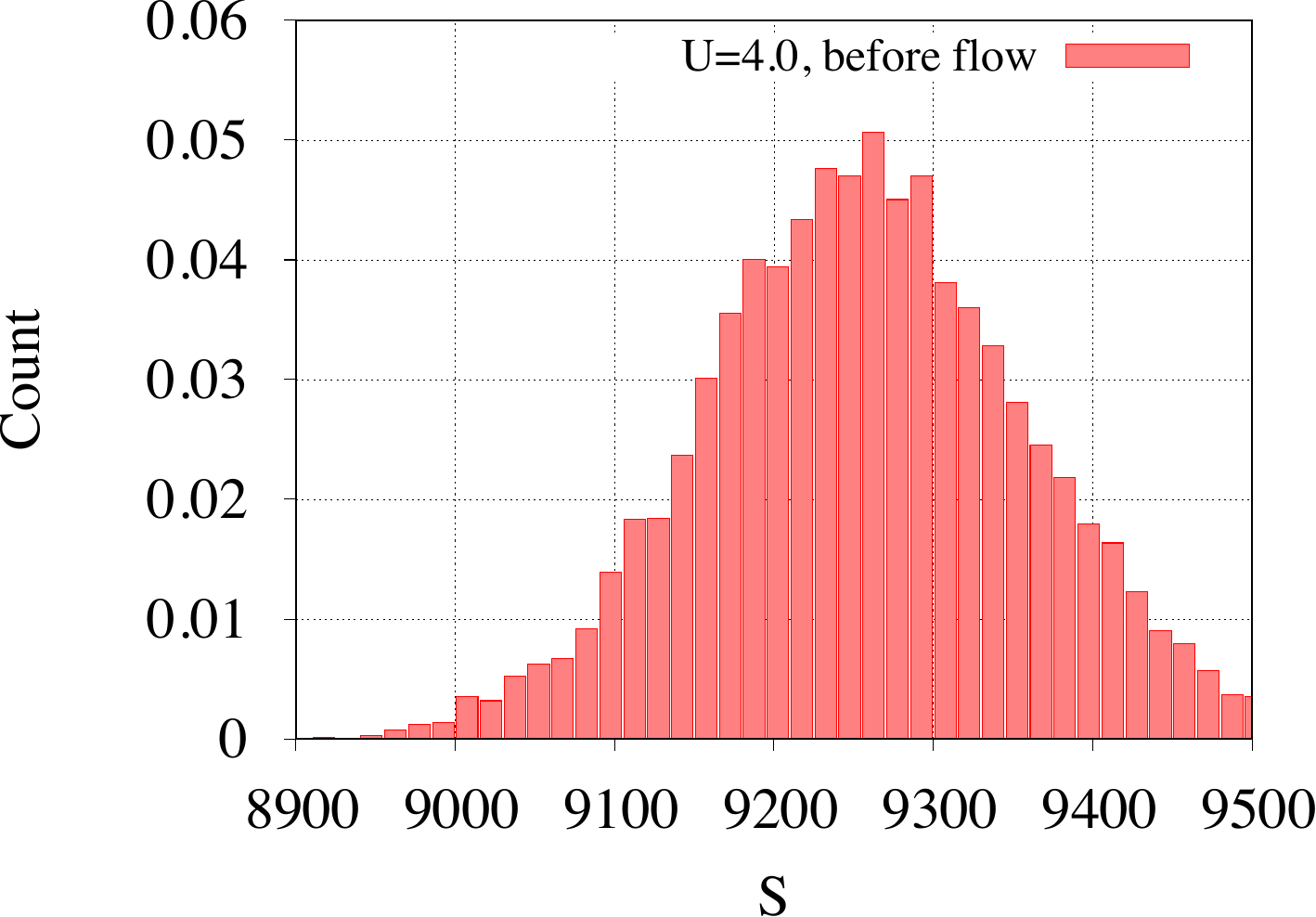}}
   \subfigure[]{\label{fig:HistogramU3After}\includegraphics[width=0.23\textwidth,clip]{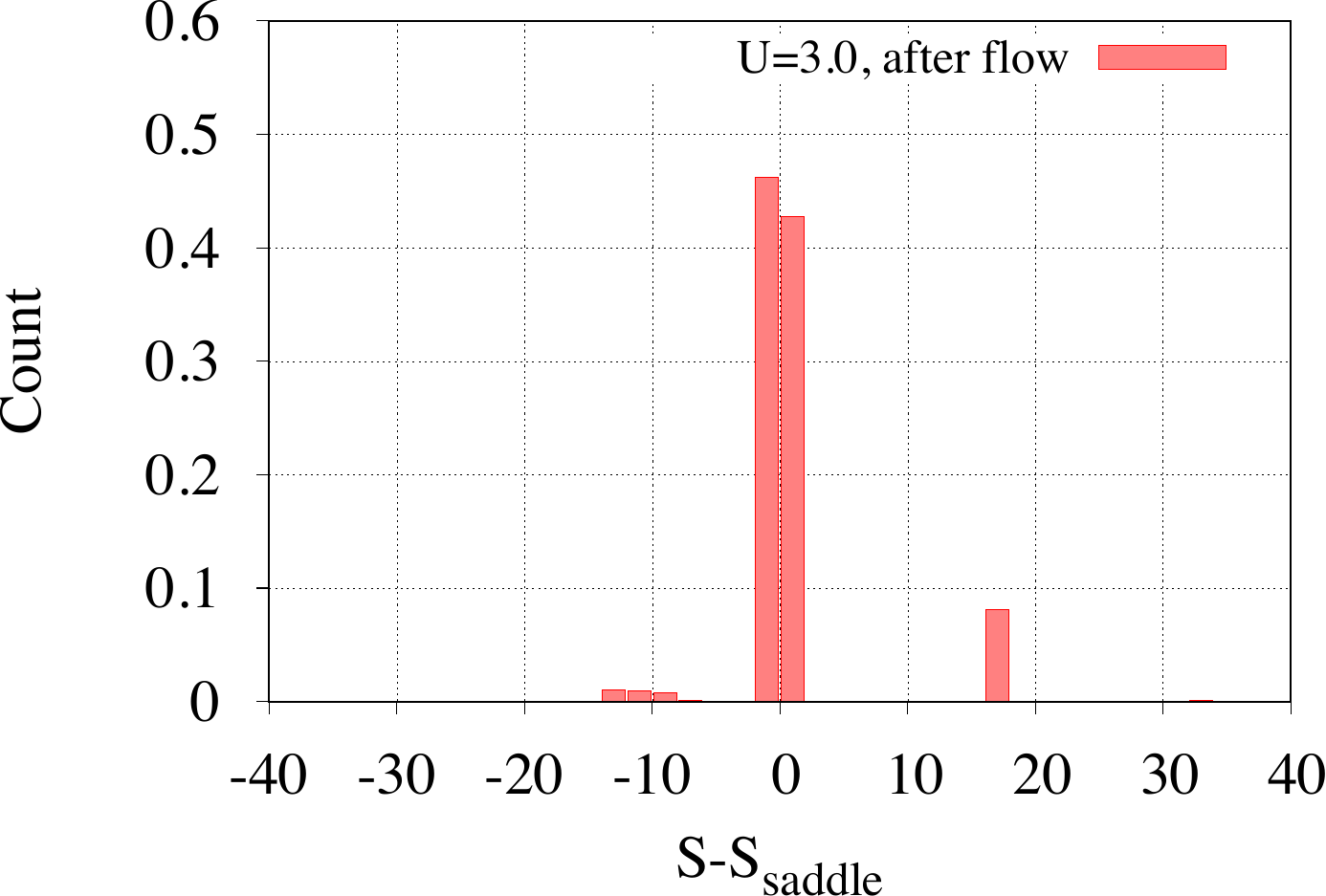}}
   \subfigure[]{\label{fig:HistogramU4After}\includegraphics[width=0.23\textwidth,clip]{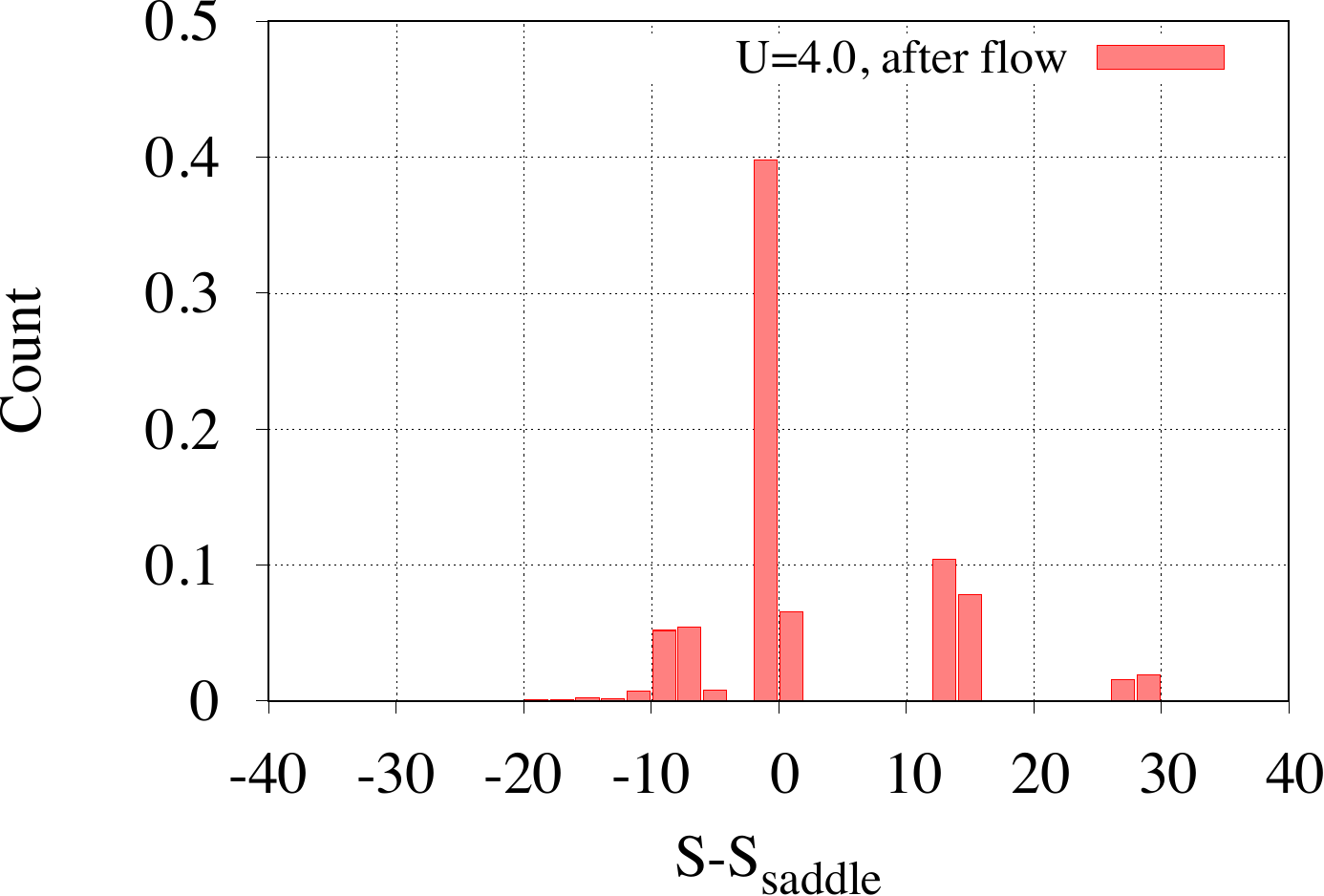}}
      \caption{ (a) and (b) Histograms for the action of configurations proposed during the Monte Carlo process, before the downward GF.  (c) and (d) Actions of the same configurations after downward GF. Calculations were made at $6\times6$ lattice with $\beta=20$ and $N_t=256$. Interactions strengths are $U=3.0$ (figures (a) and (c)) and $U=4.0$ (figures (b) and (d)). $S_{saddle}$ is the action of the desired dominant saddle points.}
   \label{fig:HistogramsMC}
\end{figure}

The important feature of the thimbles formalism is that by construction the imaginary part of the action 
is constant along the thimble. Assuming that we only sum over  the  relevant saddle points with 
$k_\sigma=1$, we can write the final form of the partition function in the thimbles formalism as
\begin{eqnarray}
\mathcal{Z}=\sum_\sigma e^{-i \operatorname{Im} S (\Phi_\sigma)} \mathcal{Z}_\sigma, \label{eq:int_thimbles_sum} \\
\mathcal{Z}_\sigma = \int_{\mathcal{I_\sigma}} D \Phi e^{- \operatorname{Re} S(\Phi) \nonumber
}. 
\end{eqnarray}

Particular saddle points can violate certain symmetries of the Hamiltonian. We would like to show that if one  constrains  integration over the set of  thimbles that due to  symmetry  considerations share the very  same  saddle  point  value of  the action then we  obtain a symmetry consistent  approximation.    First of all, the  SU(2)  spin symmetry   is  automatically   satisfied  since  our  Hubbard Stratonovich transformation  does  not  break it:  each  field  configuration   enjoys  SU(2)  spin symmetry. Thus spatial and Euclidean times translations, point group and  time  reversal  symmetries remain. As we'll show later, all these symmetries are broken  for  a  specific  field  configuration,  however, our sampling over dominant thimbles restores  them all. Spatial and Euclidean times translations, as well as point group symmetry are considered in the next section. Here we concentrate on the time  reversal  symmetry and consider not only half-filling, but also the case with finite chemical potential. 

Time reversal symmetry operator $\hat{T}=-i \sigma_2 K$ acts on spin indexes. Since the electrons and holes are connected with spin-up and spin-down electrons via the relation $\hat{a}_{\ve{x}}=\hat c_{\ve{x},\uparrow}$,  $\hat{b}^\dag_{\ve{x}}=\pm\hat c_{\ve{x},\downarrow}$ (the sign depends on sublattice), the charge can be written as $\hat q_{\ve{x}}=\hat c^\dag_{\ve{x},\uparrow} \hat c_{\ve{x},\uparrow} + \hat c^\dag_{\ve{x},\downarrow} \hat c_{\ve{x},\downarrow}-1$ and is not affected by the time reversal operator. In addition to that, $\hat T$ includes complex conjugation of the $\phi$ field. Consequently, we can write down the transformation of the Hamiltonian  \ref{eq:Z_general_cont} as:
\begin{eqnarray}
   \hat{T}^{-1} \hat H(\hat{a}^{\dagger},\hat{a}^{},\hat{b}^{\dagger},\hat{b}^{}, \phi(\tau) )  \hat{T}^{} =  \overline{ \hat H(\hat{a}^{\dagger},\hat{a}^{},\hat{b}^{\dagger},\hat{b}^{}, \phi (\tau))} = \nonumber \\ 
   \hat H (\hat{a}^{\dagger},\hat{a}^{},\hat{b}^{\dagger},\hat{b}^{}, -\operatorname{Re} \phi (\tau) + i \operatorname{Im} \phi (\tau)) .
   \label{eq:action_symm}
\end{eqnarray}
Thus, after integrating out the fermionic fields in \ref{eq:Z_general} we get action \ref{eq:action} where the following replacement of signs of $\operatorname{Re} \phi$ is made in fermionic operators: $D^{el.}_{2\tau} \rightarrow \diag{e^{-i \operatorname{Re}\phi_{{\ve{x}},\tau}  -\operatorname{Im}\phi_{{\ve{x}},\tau}+\mu \Delta \tau} }$, $D^{h.}_{2\tau} \rightarrow  \diag{e^{i \operatorname{Re}\phi_{{\ve{x}},\tau}  +\operatorname{Im}\phi_{{\ve{x}},\tau}-\mu \Delta \tau} }$. Thus the conservation of time reversal symmetry at the level of action is certain since the effective action \ref{eq:action} still respects the symmetry
\begin{eqnarray}
S(-\operatorname{Re} \Phi + i \operatorname{Im} \Phi)=\overline{S(\operatorname{Re} \Phi + i \operatorname{Im} \Phi)}.
   \label{eq:S_eff_symm}
\end{eqnarray}
However, after the shift of the integration contour to the complex space, this symmetry is satisfied configuration-wise (and not after the full integration over all thimbles \ref{eq:int_thimbles_sum} is carried out) only if both configurations $\Phi$ and $-\overline{\Phi}$ are included in the new integration contour, constructed using GF equations \ref{eq:gr_fl}.  
Using the Cauchy–Riemann conditions, the eq. \ref{eq:gr_fl} can be rewritten as
\begin{eqnarray}
\frac{d \operatorname{Re} \Phi}{dt} =  \frac{\partial \operatorname{Re} S }{\partial \operatorname{Re} \Phi}\nonumber \\
\frac{d \operatorname{Im} \Phi}{dt} =  \frac{\partial \operatorname{Re} S }{\partial \operatorname{Im} \Phi}.
  \label{eq:gr_fl_components}
\end{eqnarray}
After the replacement $ \operatorname{Re} \Phi \rightarrow -  \operatorname{Re} \Phi$, $ \operatorname{Im} \Phi \rightarrow   \operatorname{Im} \Phi$ and taking into account the symmetry \ref{eq:S_eff_symm}, we can conclude that $\frac{d \operatorname{Re} \Phi}{dt}$ changes sign, but $\frac{d \operatorname{Im} \Phi}{dt}$ remains the same. Thus for each thimble emanating from saddle point $\Phi_\sigma$ we have its counterpart attached to saddle point $-\overline{\Phi_\sigma}$, and using the relation $\Phi\rightarrow - \overline{\Phi}$, we can switch between these two thimbles.

Thus in order to restore the time reversal symmetry, we should simply automatically take into account all symmetrical (with equal $\operatorname{Re}S$ but opposite $\operatorname{Im} S$ ) configurations, which differ only by the sign of $\operatorname{Re} \Phi$. 
This restoration of the time reversal symmetry will be employed in all subsequent simulations.

In general, the expression \ref{eq:int_thimbles_sum} does not lead to a smaller sign problem, despite the elimination of the fluctuations 
of the imaginary part of the action under the integral. The problem is that many thimbles with different $\mbox{Im} S$ can participate in 
the sum, such that  cancellation effects can still be severe.  A  secondary effect is the fluctuation of the complex measure in the integral 
over the thimble  since it is in general a curved manifold in complex space. It was shown \cite{Alexandru:2016ejd, PhysRevD.101.014508}
that these fluctuations are not significant, at least for relatively small systems considered up to now.

It is impossible to predict the dominant thimble before performing  simulations for a generic model. However,
 in the particular case of the Hubbard model with charge coupled auxiliary field, there is an approximation, which can predict 
 the dominant thimbles quite precisely without any Monte Carlo input, relying solely on the parameters of the Hamiltonian \ref{eq:ham} and 
 inverse temperature $\beta$. This is the instanton gas approximation discussed in  Ref.~\cite{PhysRevB.107.045143}. Since we know the 
 dominant thimbles in advance we  can  constrain the  Monte Carlo  sampling to them thereby  neglecting other thimbles in the sum 
 \ref{eq:int_thimbles_sum}. As we will show later, this approach 
  completely eliminates the fluctuations of $\mbox{Im} S$, thus it also frees us from the necessity to switch between thimbles separated by the domain walls formed by the zeros of fermionic determinants during the generation of the field configurations in the  Monte Carlo process. This can substantially ease the ergodicity issues, which the   formulations with continuous fields are prone to \cite{Assaad_complex,Buividovich:2018yar}.  

In this paper we concentrate on the half-filling case and confront the results of simulations over the dominant 
thimbles  to  results of full QMC simulations. Since at  half-filling all thimbles are confined within the real subspace, 
we are also free from difficulties originating  from the need to generate the field configuration confined within the 
curved manifold in complex space. This  greatly  facilitates  the  computations,  and  provides an  excellent  proof  of  principle 
for the   dominant  thimbles approach.

Our tests were carried out  for the Hubbard model on hexagonal lattice. It was shown in \cite{PhysRevB.107.045143}
that the instanton gas approximation is also fully valid for the square lattice Hubbard model.
 Thus we expect that the results of our calculations are also applicable to the  square lattice.

\subsection{\label{subsec:instanton_gas}Instanton gas approximation and the role of zeros of the fermionic determinant}

Figure \ref{fig:HistogramExample} shows an example of the histogram demonstrating the relative importance of various saddle 
points for the Hubbard model. We start from the field configurations generated according to \ref{eq:z_action} in a normal
 Quantum Monte Carlo process using  the Hybrid Monte Carlo algorithm \cite{Duane87,Scalettar87,Assaad_complex,Buividovich:2018yar}. Each configuration subsequently serves as a starting point 
 for the inverse GF evolution (same as \ref{eq:gr_fl}, but with negative sign on the right hand side). At half filling thimbles 
 are confined within the real subspace, thus this flow ends in the local minimum of the action. In the complex plane, this local minimum 
 is the saddle point, that defines the thimble, hence the  initial  configuration    belongs to this  thimble. 
The histogram of the action of these final saddle point field configurations shows the relative importance of different 
thimbles within the initial integral for the partition function \ref{eq:z_action}.

As one can see, the distribution of the actions of saddle points is discrete, which corresponds to integer number of localized objects within each saddle point. These objects we identified as instantons and anti-instantons \cite{PhysRevB.107.045143}. The only difference between them is the sign of $\operatorname{Re} \Phi$ (with $\operatorname{Im} \Phi=0$ at half filling). The saddle point $\Phi_\sigma$ is now defined through the number of instantons and anti-instantons ($N_+$ and $N_-$ respectively) and their coordinates in space $\ve{X}$ and Euclidean time $T$: 
\begin{eqnarray}
    \sigma=\{(\ve{X}_1, T_1 )...(\ve{X}_{N_+}, T_{N_+}), (\ve{X}_{N_+ +1}, T_{N_+ +1})... \nonumber \\ (\ve{X}_{N_+ + N_-}, T_{N_+ + N_-}) \}. 
    \label{eq:sigma_index_full}
\end{eqnarray}
Instantons are interacting only weakly \cite{PhysRevB.107.045143}, thus the actions of the saddle points can be approximated as 
\begin{equation}
S_{N_+, N_-}=S_0 + N_+ S^{(1)} + N_- \bar{S}^{(1)},
 \label{eq:action_instantons}
\end{equation}
where $N_+$ and $N_-$ are number of instantons and anti-instantons respectively, $S_0$ is action of vacuum saddle (corresponds to zero fields $\Phi=0$) and $S^{(1)}$ is action of one instanton. Contributions of instanton and anti-instanton in the action are equal at half filling since the action is real $\operatorname{Im} S^{(1)}=0$. Due to translational symmetry of the action and localized nature of (anti)instantons, their actions are also independent on their coordinates $(\ve{X}_i, T_i)$.  Thus all distribution show us the total number of instantons and anti-instantons in the saddle point $N_+ + N_-$, and each peak in  Fig.~\ref{fig:HistogramExample} is in fact a collection of equivalent saddles with fixed $N_+ + N_-$ but different individual numbers of (anti)instantons and coordinates of these objects. The maximum of the distribution in Fig.~\ref{fig:HistogramExample} gives 
us the instanton density in the dominant saddles. This density can be predicted even without plotting such histograms, using the model of 
instantons as a  weakly interacting gas in 2+1D Euclidean   volume  \cite{PhysRevB.107.045143}.  Comparison of this prediction with real QMC results is shown in Fig.~\ref{fig:InstantonGasConstruction}. 
As one can see, the prediction is reasonably accurate.    

Unfortunately, this prediction is not enough to reproduce fully the  physics of the Hubbard model. 
As was shown in \cite{PhysRevB.107.045143}, the pure saddle point approximation corresponds to increasingly localized, but disordered spins. 
In  fact,  with this  saddle point  approximation,  it  was not possible to reproduce the Gross-Neveu transition and appearance of the gap
 at the Dirac point, which  for  the  Hubbard model on  honeycomb lattice  occurs at  
  $U \simeq 3.8$ \cite{PhysRevX.3.031010}.

Let us investigate more deeply the reasons behind this deficiency. Fig.~\ref{fig:actions} shows the profile of the action~\ref{eq:action}  
along a  two-dimensional  cross-section of the configuration space defined as  follows.   Consider   two field  
configurations  $\phi^{(1)}_{{\ve{x}},\tau}$   and $\phi^{(2)}_{{\ve{x}},\tau}$    corresponding  to  single   instantons  at  different  
locations. The  point $(x,y)$ in Fig.~\ref{fig:actions}    corresponds  to the  field  configuration    $\phi =  x \phi^{(1)} + y\phi^{(2)} $
such  that $(0,0) $  corresponds  to the vacuum,  $(1,1)$  to a   two instanton  saddle  point and $(1,0)$ and $(0,1)$  to single instanton saddles.
Due to the translational invariance, the action is the same at two latter   saddles.
One can clearly see the domain walls separating the saddles. They are formed by configurations, where the action goes to $+\infty$. 
These domain walls are nothing else but zeros of fermionic determinants. Naively, the weight of these configurations tends to zero within 
the integral for the partition function \ref{eq:Z_general}. However, their real role is enhanced by the fact that at least
some observables diverge at these points. This phenomenon was studied in details in \cite{PhysRevE.106.025318}
where it was argued that these domain walls are responsible for the fat tails of the distributions of the fermionic observables. 
Here we only briefly repeat the reasoning behind this fact, while the interested reader can look into the paper \cite{PhysRevE.106.025318}
for mathematical details. The zeros of the fermionic determinant corresponds to the situation where the  corresponding fermionic operator has a zero 
eigenvalue. Typical fermionic observable computed through the Wick theorem includes some combination of the inverse fermionic operators
 (fermionic propagators). Due the presence of zero eigenvalues, the latter diverge once we approach the domain walls between thimbles.
  In order to reveal the influence of the domain walls onto the long range spin order, we plot the ratio of long range spin-spin correlations
   to short range spin-spin correlations for the same 2D cross section in  configuration space. Generally, the spin-spin correlator for 
   sites within the same sublattice is computed as an average over configurations $\Phi$, generated with the weight \ref{eq:z_action}:
 \begin{eqnarray}
\langle \hat {\vec S}_{\ve{x}} \hat {\vec S}_{\ve{x} + \ve{r}} \rangle = \lim_{N_{conf.} \rightarrow +\infty} 
\frac{1}{N_{conf.}}\sum_\Phi C_{\ve{x}, \ve{x} + \ve{r}} (\Phi),
\label{eq:spin_correlator}
\end{eqnarray}
 where $N_{conf.}$ is  the  number of generated configurations and
 \begin{eqnarray}
 C_{\ve{x}, \ve{x} + \ve{r}} (\Phi) = \frac{3}{4} \left( |g_{\ve{x}, \ve{x} + \ve{r}}(\Phi)|^2  +  |g_{\ve{x} + \ve{r}, \ve{x}}(\Phi)|^2 \right).
 \label{eq:CSPhi} 
\end{eqnarray}
$g_{\ve{x}, \ve{x} + \ve{r}}(\Phi)$ is the  equal time fermionic propagator for electrons computed for 
the configuration of auxiliary fields $\Phi$. We plot the ratio of observables 
$C_{\ve{x}, \ve{x} + \ve{r}} (\Phi)$ for individual configurations
 \begin{eqnarray}
R(\Phi)=\frac{C_{\ve{x}, \ve{x} + \ve{L}} (\Phi)}{C_{\ve{x}, \ve{x} + \ve{r_0}} (\Phi)}.
\label{eq:spin_ratio}
\end{eqnarray}
Here $\ve{L}$ corresponds to maximum  spatial distance  on the lattice and $\ve{r_0}$ corresponds to 
the distance to the  next-to-nearest neighbour lattice site.

The ratio $R(\Phi)$ is plotted on Fig.~\ref{fig:spins}. The ratio is negligibly small for the configurations near the saddle points, 
but quickly increases once we approach the domain walls formed by the zeros of the fermionic determinants. 
Now the deficiency of the saddle point approximation is evident: in fact, for the particular case of the charge coupled auxiliary fields 
the configurations away of the saddle points are responsible for the long range spin order.

Many  question now arises. What is the  quality of the approximation in  which  we  constrain  the  Monte Carlo sampling 
to  the  dominant  thimbles?   As opposed to the  instanton  gas  approximation,  can it    account for  symmetry  breaking 
mass  generation?

\section{\label{sec:formalism_QMC}QMC on a limited set of thimbles}
Our aim is to construct a  Monte Carlo algorithm, which will generate the auxiliary field configurations according to 
the distribution $e^{-S(\Phi)}$   in  the  restricted  space of    thimbles, $\mathcal{I}_\sigma, \sigma\in D$ attached to dominant saddle points. The corresponding
 partition function is written as:
\begin{eqnarray}
\mathcal{Z}_D=\sum_{\sigma \in D}\int_{\mathcal{I}_\sigma} D \Phi e^{-S(\Phi)}.
\label{eq:Z_thimble}
\end{eqnarray}

In order to restrict the Monte Carlo process within the designated thimbles, we modify the generation of auxiliary field configurations.
First,  the proposal of the new configuration is made in the same manner as it is done in the  Hybrid Monte Carlo (HMC) technique: 
via artificial Hamiltonian dynamics for potential equal to the action \ref{eq:action} \cite{Duane87,Scalettar87,Assaad_complex}.
Ideal Hamiltonian dynamics cannot cross the domain wall of infinite potential. However,  an  instanton-anti-instanton pair can be created 
without crossing the domain wall \cite{PhysRevB.107.045143}.  In addition to this effect, the  numerical Hamiltonian dynamics can 
occasionally cross the domain wall due to the  finite MD time step. Thus the usual Metropolis accept-reject step from HMC algorithm should 
be supplemented by an additional check that the proposed new configuration still belongs to the designated thimble. This check is done in 
the same manner as the construction of the histogram \ref{fig:HistogramExample}. We perform the inverse GF according to the equation 
\ref{eq:gr_fl} with minus sign on the right hand side and starting from the proposed configuration. This gradient flow ends up in the
saddle point for the thimble, which the proposed configuration belongs to. 
We check that this saddle point at the end of the flow has still the same number of instantons and anti-instantons as the desired saddle, 
and accept the proposed configuration only if this is true. An illustration of this process is shown in the figure 
\ref{fig:HistogramsMC} for two simulations for $U=3.0$ and $U=4.0$. We plot histograms of the actions of the proposed configuration 
alongside with the histograms of the actions of the configurations after the downward GF. The former histograms show broad and featureless 
distributions, while the latter show  discrete distributions similar to fig. \ref{fig:HistogramExample}. It gives us a quite simple criterion 
for the additional accept/reject step of the Monte Carlo algorithm: only those proposed configurations can be accepted, where the action after the  inverse 
GF falls in the peak corresponding to the desired saddle.

In practice we start from the single dominant saddle point field configuration predicted by the instanton gas model, which gives us the number of instantons $N_+$ and anti-instantons $N_-$ and their coordinates $(X_i, T_i)$. $N_+=N_-$, since the most probable configuration must have equal number of instantons and anti-instantons for obvious combinatorial reasons \cite{PhysRevB.107.045143}. After the MC update and downwards GF we only check that we returned back to the saddle with the same action and $N_+=N_-$ as the starting field configuration. Since we do not control the coordinates of individual (anti)instantons, they are allowed to move freely through the lattice. HMC can facilitate this move, since it is possible to change the spatial location of the instanton by tunneling through the local peak of the action. One can observe that it is possible to construct a trajectory connecting points $(1,0)$ and $(0,1)$ in Fig.~\ref{fig:actions} without crossing the domain wall of zeros of determinant. Since these points on the figure correspond to the instantons at different spatial locations, free movement between such points restores spatial translational and rotational symmetries. Movement along the Euclidean time axis is even simpler since it corresponds to the continuous symmetry (in the limit $\Delta \tau=0$), thus there should be no domain walls between field configurations $\{ \phi_{x, \tau} \}$ and $\{ \tilde\phi_{x, \tau} \}=\{ \phi_{x, \tau+1} \}$, at least if $\Delta \tau$ is small enough. In the saddle point approximation, this movement corresponds to the zero mode of Hessian of the saddle point, discussed extensively in \cite{PhysRevB.107.045143}.

Algorithm constructed in this way stays indefinitely within the series of interconnected thimbles with fixed $N_+$ and $N_-$ numbers. This thimbles are attached to the dominant saddle points, thus giving us the desired approximation to 
 the integral for the partition function \ref{eq:Z_thimble}, where $D$ is a set of all possible saddle points with fixed $N_+$ and $N_-$, but different coordinates of individual (anti)instantons. We stress that despite formally multiple thimbles participate in the partition function, the fluctuations of $\operatorname{Im} S$ would still be absent even  away of half filling, since the $\operatorname{Im} S$ is fully defined by $N_+$ and $N_-$ numbers \ref{eq:action_instantons}.

Configurations generates with this algorithm are subsequently used for the calculation of the fermionic observables. First we look at charge and spin order
 parameters. Following 
\cite{Buividovich:2018yar} we compute squared spin per sublattice
\begin{eqnarray}
\langle \hat {\vec S}^2  \rangle = \left\langle  \left( \sum_{\ve{x}\in A} \hat {\vec S}_{\ve{x}}  \right)^2 \right\rangle
\label{eq:sq_spin}
\end{eqnarray}
and squared charge per sublattice
\begin{eqnarray}
\langle \hat {Q}^2  \rangle = \left\langle  \left( \sum_{\ve{x}\in A} \hat q_{\ve{x}}  \right)^2 \right\rangle ,
\label{eq:sq_charge}
\end{eqnarray}
where $A$ denotes one of the two sublattices of the bipartite lattice.
These observables can be rewritten in terms of averages over configurations of auxiliary fields generated according to the 
distributions in \ref{eq:z_action} or \ref{eq:Z_thimble}:
\begin{eqnarray}
 \langle \hat {\vec S}^2  \rangle =  \lim_{N_{conf.}   \rightarrow +\infty} \frac{3}{4 N_{conf.}} \sum_\Phi 
\left( \sum_{\ve{x},\ve{y}\in A, \ve{x}\neq\ve{y}}  C_{\ve{x}, \ve{y}} (\Phi)  + \right.  \nonumber  \\     \left. \sum_{\ve{x}\in A} \left( 1 - 2 \operatorname{Re}  g_{\ve{x}, \ve{x}}(\Phi) + 
2 |g_{\ve{y}, \ve{x}}(\Phi)|^2  \right) \right)
\label{eq:sq_spin_grf}
\end{eqnarray}
\begin{widetext}
\begin{eqnarray}
\langle \hat Q^2  \rangle =   \lim_{N_{conf.} \rightarrow +\infty} \frac{2}{N_{conf.}} \sum_\Phi 
\left( \sum_{\ve{x}\in A } \left( \operatorname{Re} g_{\ve{x}, \ve{x}}(\Phi) - |g_{\ve{x}, \ve{y}}(\Phi)|^2 \right) + \right. \nonumber \\
\left. \sum_{\ve{x},\ve{y}\in A,\ve{x}\neq\ve{y} } \operatorname{Re}  \left( g_{\ve{x}, \ve{x}}(\Phi) g_{\ve{y}, \ve{y}}(\Phi)  -
  g_{\ve{x}, \ve{y}}(\Phi) g_{\ve{y}, \ve{x}}(\Phi)  -
   g_{\ve{x}, \ve{x}}(\Phi) g^*_{\ve{y}, \ve{y}}(\Phi) \right)  \right)
\label{eq:sq_charge_grf}
\end{eqnarray}
\end{widetext}

Real parts of the expression are taken in order to account for trivial symmetry of the action \ref{eq:action} with respect to the exchange of the sign of real part of the $\Phi$ field. It means that for each configuration, there is a counterpart with exactly the same weight in the partition function but opposite real parts of the fields, which corresponds to the exchange of $g$ and $\bar g$ in all fermionic observables. According to the discussion concerning the time reversal symmetry in the section \ref{subsec:formalism_intro}, it corresponds to the enhanced estimator restoring the time-reversal symmetry. 

For the benchmark  QMC calculation, we use  field configurations $\Phi$ generated according to the distribution in \ref{eq:z_action}, 
whereas  for the  dominant thimbles approximation, the distribution of  \ref{eq:Z_thimble} is sampled. The complete simulation spanning 
over all thimbles can be done via the addition of the small second auxiliary field coupled to spin density. It helps the 
Hybrid Mote Carlo to to penetrate through the domain walls by increasing the overall dimensionality of the configuration 
space \cite{Assaad_complex,Buividovich:2018yar}.

\begin{figure}[]
   \centering
   \subfigure[]{\label{fig:ResultsSpin}\includegraphics[width=0.35\textwidth,clip]{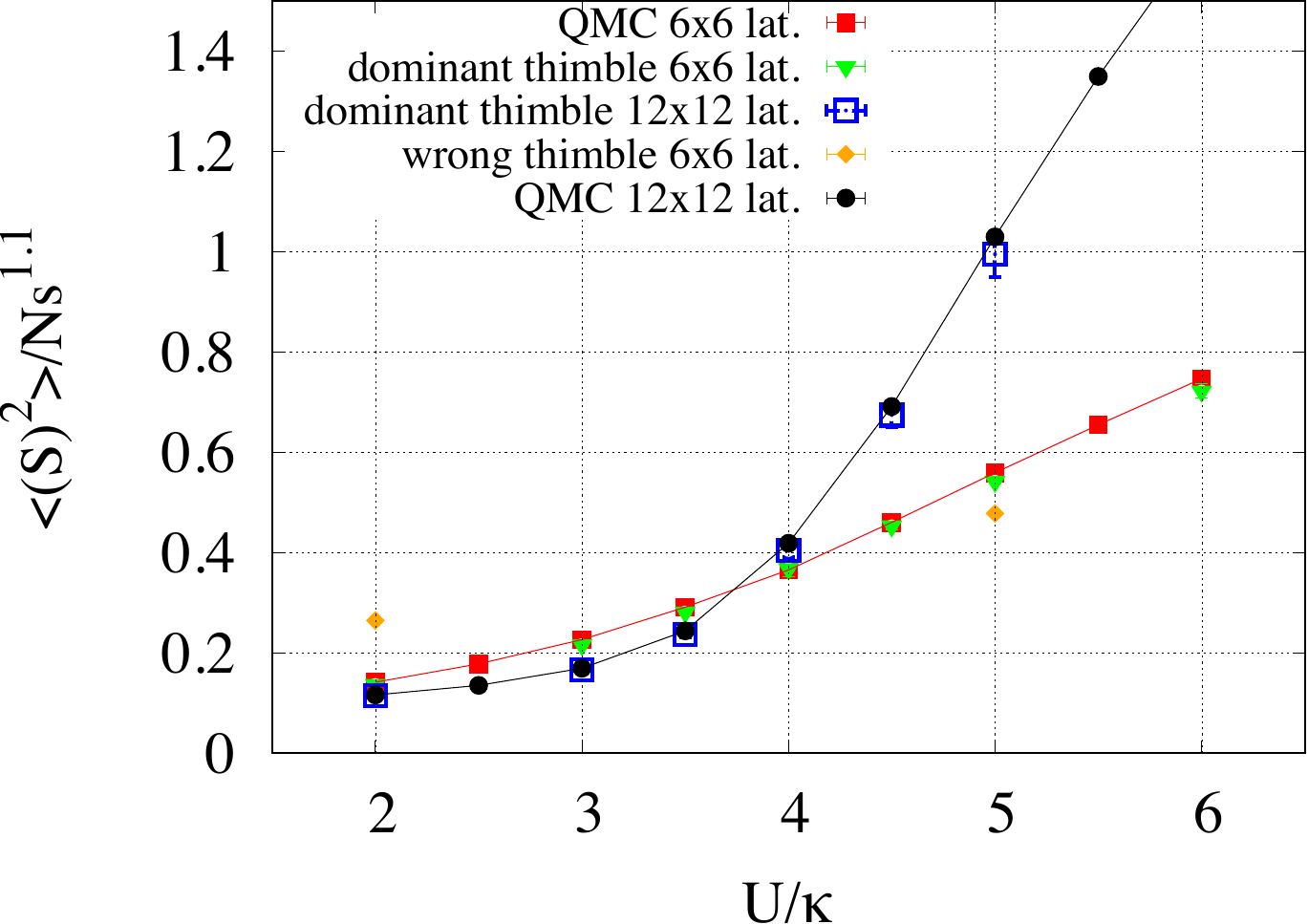}}
   \subfigure[]{\label{fig:ResultsCharge}\includegraphics[width=0.35\textwidth,clip]{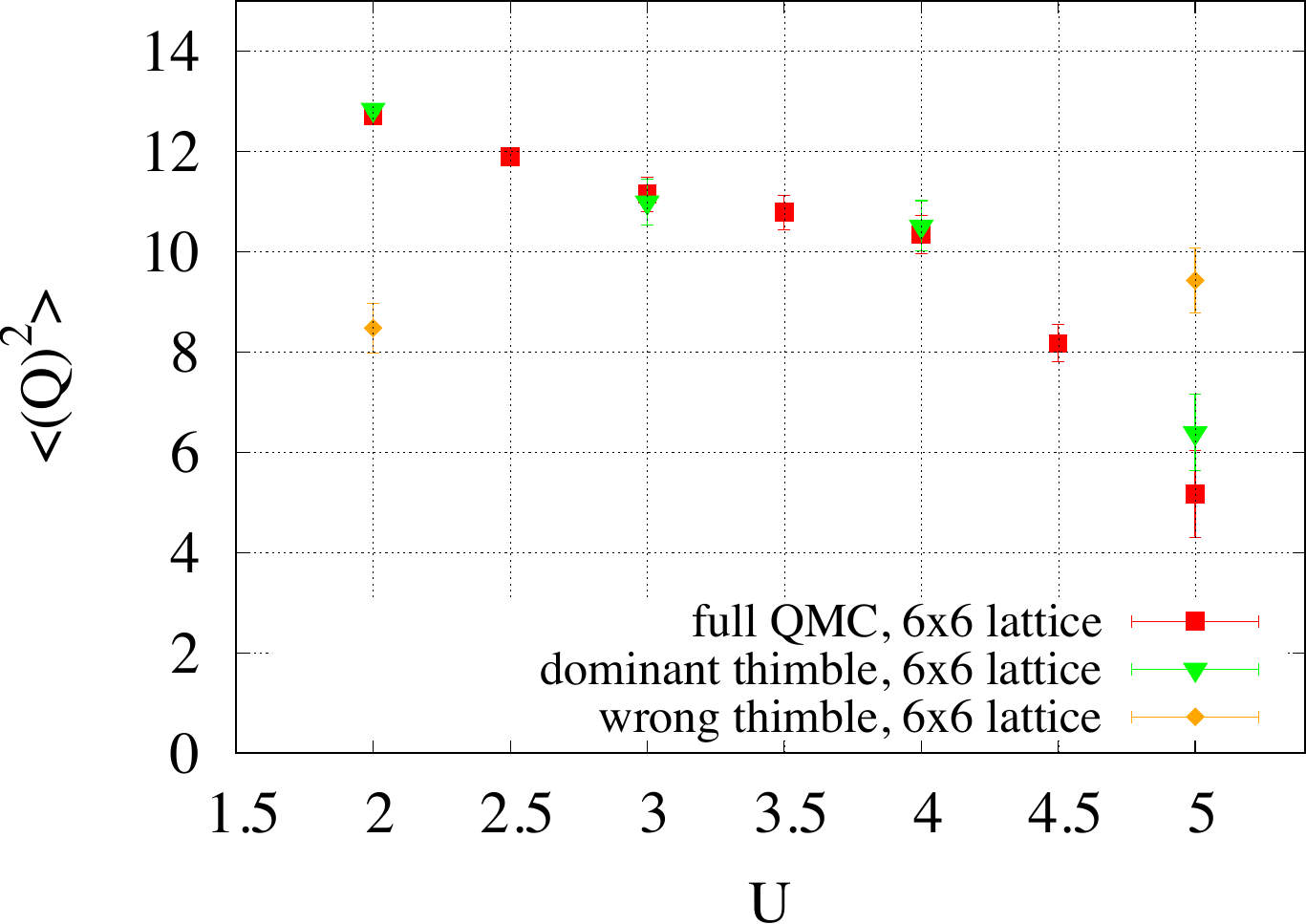}}
   \caption{Squared spin (a) and squared charge (b) on sublattice computed using full QMC  \ref{eq:z_action} and with Monte Carlo process constrined to a  single  thimble \ref{eq:Z_thimble}. Rescaling is done for the squared spin observable using   critical exponents of  the  first order  $\epsilon$-expansion,  $\frac{\langle \vec{S}^2 \rangle }{N_s^2} = N_s^{-\beta/\nu} F\left( L^{1/{\nu}} \left(U - U_c \right) \right)$.    Here we omit  correction to  scaling.
           The dominant saddle points are always  predicted using instanton gas model 
           with hardcore repulsion. The \textit{wrong thimble} results  at $U=2.0$   refer  to  the one attached to the 40-instanton saddle on a 
           $6\times6$ lattice (the true dominant saddle is the vacuum one in this case); wrong thimble at $U=5.0$ means the one attached 
           to vacuum, while true dominant saddles contain 25 (anti)instantons. Calculations were done on $6\times6$ and $12\times12$ 
           lattices with $\beta=20$ and $N_t=256$.}
   \label{fig:ResultsComparison}
\end{figure}

Benchmarks   are shown in Figs.~\ref{fig:ResultsSpin} and \ref{fig:ResultsCharge}. In addition to the dominant thimbles results, 
we also show the results of the simulation over the thimble attached to the wrong saddle point: many instanton saddle at small 
$U$ and vacuum at large $U$. The  squared spin results are rescaled using   critical exponents  corresponding to the  first order  $\epsilon$-expansion,  $\beta/\nu=0.9$ (see Ref.~\cite{AssaadHerbut2013} for the accuracy of this approximation).
With this  rescaling  the   critical point  corresponds  to the coupling constant at  which the  scaled  square  spin  is independent on the system size.
\footnote{We  omit  corrections to  scaling. }

 \begin{figure}[]
   \centering
   \subfigure{\includegraphics[width=0.35\textwidth,clip]{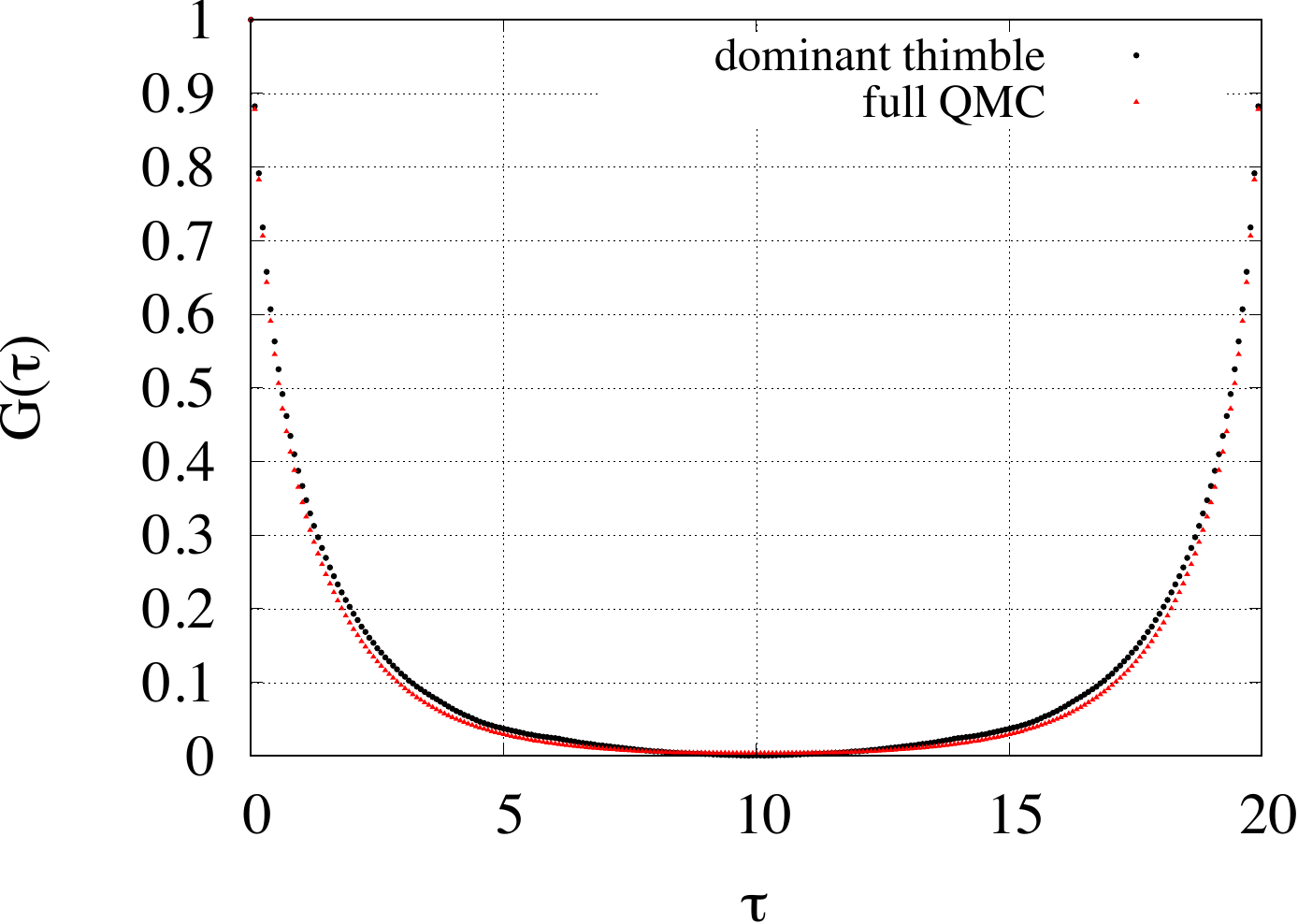}}
   \subfigure{\includegraphics[width=0.35\textwidth,clip]{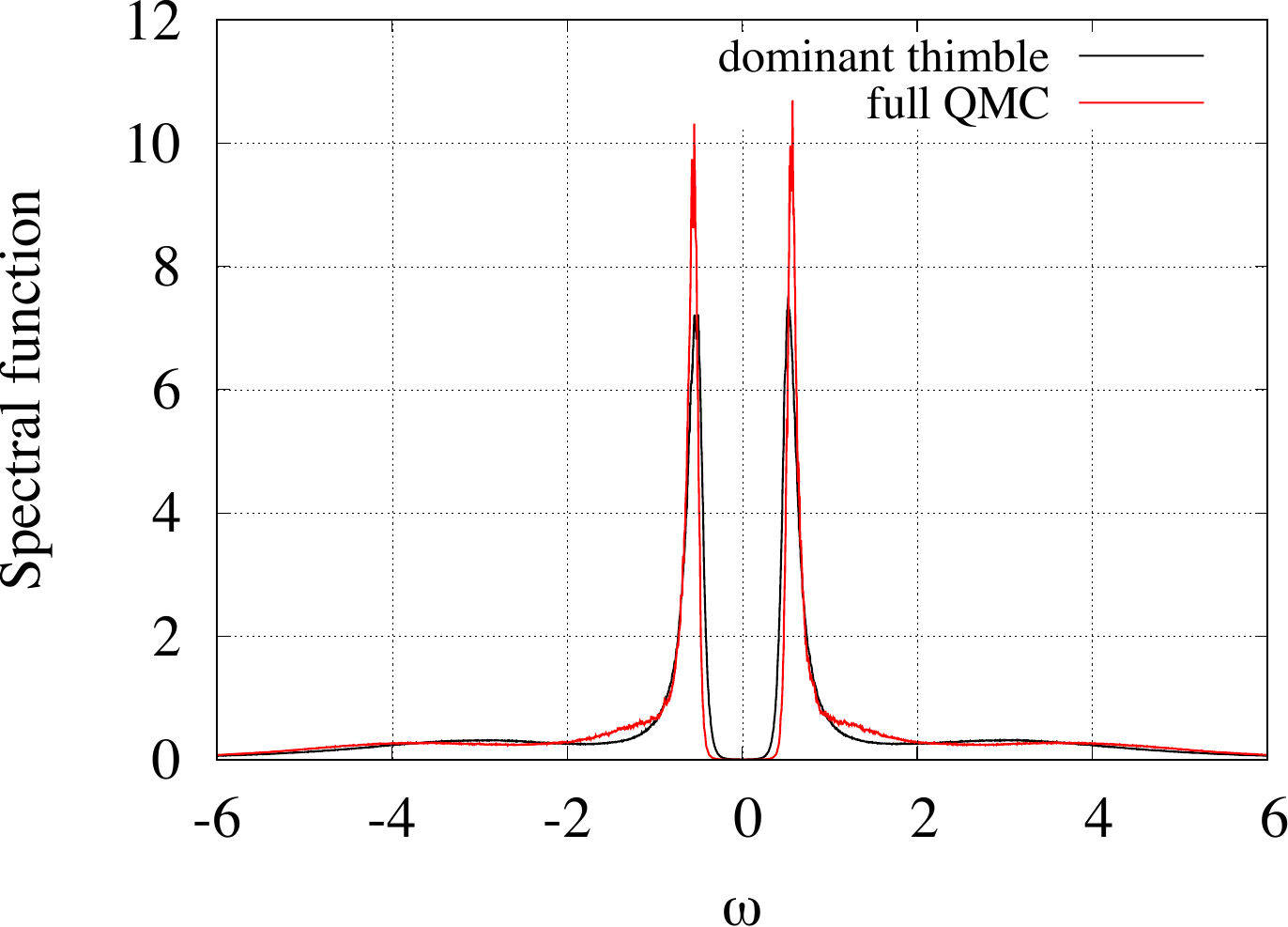}}
      \caption{Single-particle Euclidean time propagators (a) and corresponding spectral functions (b) at the Dirac point obtained in full QMC calculation and in the Monte Carlo process bounded within one dominant thimbles. Calculation was done on $12\times12$ lattice for $U=5.0$ at inverse temperature $\beta=20$.}
  \label{fig:ResultsMass}
\end{figure}

As one can see, the dominant thimbles results are hardly distinguishable from the full simulation for both $6\times6$ and $12\times12$ lattices.
This implies  that we have similar  crossing  points  for the  full  and  dominant thimbles  approximation.  In Fig.~\ref{fig:ResultsCharge} we  see  that  the squared charge does not decay with increased interaction strength if we are using wrong saddle point. This is in line with the previously reported results of \cite{Buividovich:2018yar}, where the results for squared charge were reported to deviate from the correct ones due to the ergodicity issues with the single field HMC algorithm: in this case MD can not cross the domain walls between different thimbles.

As a final check,we compute the single electron spectral function in the dominant 
thimbles simulation. We compute it using Euclidean time fermionic propagator  and the  Algorithms for Lattice  Fermions (ALF)  \cite{ALF_v2} implementation
of the  Stochastic analytical  continuation \cite{Beach04a,Sandvik98} to  real  time.  
The saddle point approximation \cite{PhysRevB.107.045143} leads to 
the absence of the mass gap at the Dirac point even in the strong coupling limit. On the contrary, in the dominant thimbles approximation, 
the spectral function is hardly distinguishable from the one obtained in the full simulation. Thus we can conclude that it is indeed 
enough to only simulate the limited set of dominant thimbles in order to  obtain the Mott insulator with long range Antiferromagnetic (AFM) spin 
order at $U>4$.

\section{\label{sec:Conclusion}Conclusion}

We demonstrated that the  dominant thimbles  approximation  produces precise  results  for the  Hubbard model  at  half-filling. 
In  fact,  within our precision  the  dominant thimbles  results match precisely  with    our  benchmark  results. 
 In particular, we showed the appearance of long range spin-spin order with 
concomitant  suppression of the squared charge.  Additionally,   the single particle spectral function shows the 
appearance of the gap in the spectrum at the Dirac point  and  the  imaginary  time  Green  functions  of  the dominant 
thimbles  approximation  turns  out  to  be  very close   to our  benchmark  result.

The crucial feature of the considered dominant thimbles approximation  is a  understanding of the  saddle point  structure of the 
path  integral.  In  our  specific   half-filled  particle-hole  symmetry  case, this knowledge  stems from  the 
instanton gas  approximation discussed in  \cite{PhysRevB.107.045143}. 
 This  insight  provides a    classification  of  saddle 
points in terms  of  instantons and anti-instantons,  and it is  possible  to  predict   the  number of  instantons and  anti-instantons 
in  the dominant  saddle  points.  This information suffices  to constrain the simulation  to  a series of dominant thimbles. 
The  half-filled  case  turns  out to  be  especially  simple since  thimbles  are  constrained  to the  real  space.  Generically,  
the  thimble  is  curved  manifold  in  complex space, and it is   considerably  more  complicated  to  construct it.

As we showed above,  the  dominant thimbles  approximation  does not   break  
any  symmetries  of the Hamiltonian.  Furthermore,  we observe  a   transition  to  an  antiferromagnetic  insulating  state,   that 
certainly  seems  to be  continuous  given our  admittedly  limited  data  set.    Since  generically   symmetries and  dimensionality  
pin down  universality,  we  conjecture  that   the  quantum phase  transition  we  observe  within the dominant thimbles approximation 
belongs  to the O(3) Gross-Neveu phase  transition  \cite{Herbut09} as  observed  in the  Hubbard  model on the  honeycomb lattice 
\cite{AssaadHerbut2013,Toldin14,Otsuka16}.

The most important consequence of  our  results,  is  a strong motivation to consider similar dominant
 thimble approximations also away of half filling.  It was shown that the instanton gas approximation retains essentially the same structure 
 also at finite chemical potential \cite{Ulybyshev:2019fte}.  Thus the only  modification required  to  implement  this approach away from half-filling would 
 be to construct (approximate) thimbles in complex space. While this is still a formidable task, we will be at least freed from 
 the ergodicity  issues connected with the necessity  for the algorithm to jump between relevant thimbles with different values of $\operatorname{Im} S$.  This  is sometimes difficult and need additional care \cite{Ulybyshev:2019fte, PhysRevD.100.114510}. 
 The only input for such  an algorithm would be 
 the \textit{ab initio} prediction of the dominant thimbles for the case of non zero chemical potential from the complexified instanton 
 gas approximation, which will be the  subject of  follow up papers.

We  again stress that   even  away  from half-filling, the  dominant  thimbles   approximation  does  not   explicitly  break 
 symmetries.   At  half-filling  the approximation has the potential of   accounting for  long  ranged order  and  
 concomitant  spontaneous  symmetry   breaking.  Away  from half-filling  we  will not  suffer for  the sign problem such 
 that  irrespective on the  the quality of the approximation our   simulations  will   provide  a  \textit{symmetry}  consistent  
 point of view on  doping  an antiferromagnetic Mott insulator. 

\begin{acknowledgments}
MU  thanks  the  DFG   for financial support  under the projects UL444/2-1. 
FFA   acknowledges financial support from the DFG through the W\"urzburg-Dresden Cluster of Excellence on Complexity and Topology in Quantum 
Matter - \textit{ct.qmat} (EXC 2147, Project No.\ 390858490)   as  well as  the SFB 1170 on Topological and Correlated Electronics at
 Surfaces and Interfaces (Project No.\  258499086). Computational resources were provided by the Gauss Centre for Supercomputing e.V. 
 (www.gauss-centre.eu) through the John von Neumann Institute for Computing (NIC) on the GCS Supercomputer JUWELS~\cite{JUWELS} at 
 J\"ulich Supercomputing Centre (JSC). 

\end{acknowledgments}

\bibliography{thimbles}

\end{document}